\documentclass[11pt]{revtex4}
\usepackage{graphicx}

\def\be{\begin{equation}}
\def\ee{\end{equation}}
\def\bea{\begin{eqnarray}}
\def\eea{\end{eqnarray}}

\begin{document}
\vspace*{3cm}
\title{RADIO DETECTION OF COSMIC RAY AIR SHOWERS WITH CODALEMA}

\author{D. Ardouin$^1$, {\bf A. Bell{\'e}toile$^{1}$}, D. Charrier$^1$, R. Dallier$^1$,
 L. Denis$^3$, P. Eschstruth$^4$, T. Gousset$^1$, F. Haddad$^1$, J. Lamblin$^1$,
  P. Lautridou$^1$, A. Lecacheux$^2$, D. Monnier-Ragaigne$^4$, O. Ravel$^1$}
\address{$^1$SUBATECH, 4 rue Alfred Kastler, BP
  20722, F44307 Nantes cedex 3\\$^2$LESIA, Observatoire de Paris - Section de Meudon, 5
  place Jules Janssen, F92195 Meudon cedex\\$^3$Observatoire de Paris - Station de
   radioastronomie, F18330 Nan\c{c}ay\\$^4$LAL, Universit\'e Paris-Sud, B\^atiment 200, BP 34,
  F91898 Orsay cedex}

\begin {abstract}
Studies of the radio detection of Extensive Air Showers  is the goal of the 
demonstrative experiment CODALEMA. Previous analysis have demonstrated that detection around 
$5.10^{16}$~eV was achieved with this set-up. New results allow for the first time to study 
the topology of the electric field associated to EAS events on a event by event basis.
\end{abstract}
\maketitle

\section{Introduction}

Large experiments, like Agasa~\cite{agasa} or Hires~\cite{fly}, have brought into light
 difficulties concerning the determination of the origin and energy of the Ultra High Energy 
 Cosmic Rays (UHECR). In a near future, the giant hybrid experiment PAO~\cite{auger} could be
  in situation to enlighten this problem by furnishing a larger set of events, using both particle
   and fluorescence detectors. 

Beside this, some experiments~\cite{casa-mia}$^,$~\cite{cascade-grande}$^,$~\cite{nim-ard} have
 recently started to detect Extensive Air Showers (EAS), taking advantage of the radio emission 
 associated to their development. Actually, the radio detection was proposed in the
  1960's~\cite{ask} and gave birth to several experimental investigations~\cite{allan} but this
   effort was quickly abandoned. Thanks to technical improvements, recent results~\cite{prl-ard}
   $^,$~\cite{huege} demonstrate that EAS radio detection is now feasible. Furthermore, advantages
    of this method (100 \% duty cycle, low cost of the detector, etc...) make it a promising tool
     for future detectors. 
The CODALEMA experiment is a part of this demonstrative effort. Despite measurement of UHECR 
around $10^{20}$ eV is the admitted goal of such a method, a proof-principle demonstration at this
 energy would suffer of a lack of statistic. This can be avoided by working around $10^{17}$ eV
  with a much lower signal amplitude, specially far from the shower core. Nevertheless, this
   signal should remain measurable~\cite{huege}. 
    Considering a vertical shower
    falling upon the detector, the predicted transient should reach 150 $\mu$V/m with a 10 ns
     FWHM duration~\cite{casa-mia}$^,$~\cite{nim-ard}. Transposed in the frequency domain, the 
     corresponding pulse spectrum should extend from 1 to 100 MHz. Thus, a broad frequency
      bandwidth antenna should permit to recover the original pulse shape, allowing to determine
       the energy and nature of the primary with minimal assumptions concerning its 
       electromagnetic shower signature.

\section{Principle of the experiment}

\begin{figure}
\begin{center}
\includegraphics[height=8.cm]{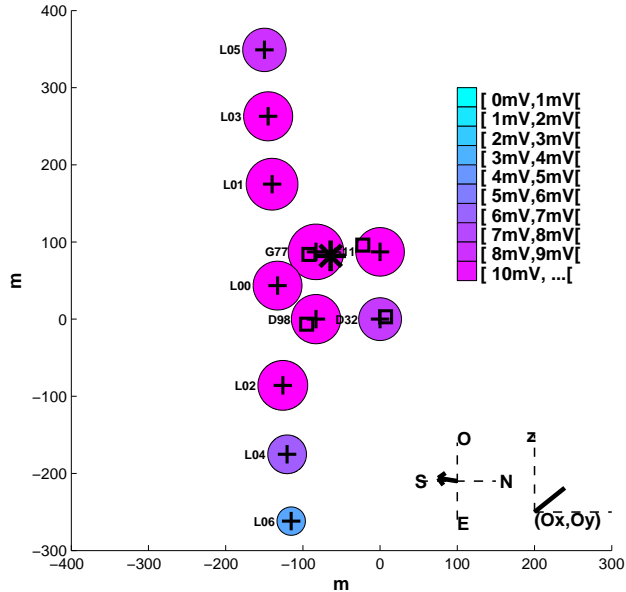}
\end{center}
\caption{Display of the CODALEMA set-up with a superposed transient EAS event. The squares
 indicate the location of the particle stations. 
Each cross corresponds to one antenna and the size and color of the circles superposed 
are proportional
 to the measured voltage. The arrival direction and elevation angle of the event are mentioned.
  The star indicates the reconstructed shower core location from the radio signal. }
\label{fig:chp_fh}
\end{figure}

The CODALEMA experiment~\cite{rav04} takes place at the radio observatory of Nan\c cay (France).
 The present related setup (see Fig.~\ref{fig:chp_fh}) uses 11 log-spiral antennas (originally 
 part of the DecAMetric array~\cite{DAM}) and 4 particle scintillator stations acting as a trigger
  (originally designed as prototypes for the PAO array~\cite{boratav}). 

Broadband antenna signals (frequency bandwidth of 1-100 MHz) are amplified (35 dB) and recorded 
in a waveform mode (8 bits ADC, 500 MHz sampling frequency, 10 ${\mu}$s recording time) using a 
common time reference given by the trigger. Due to sensitivity considerations with those ADCs, 
antenna signals are band-pass filtered (24-82 MHz). 
Particle stations are made up of two $2.3$ m$^2$ layers of acrylic scintillator, each being read out by a 
photomultiplier. Signals from upper layers of each station are digitized (8 bits ADC, 100 MHz
 sampling frequency, 10 ${\mu}$s recording time). The coincidence between top and bottom layers
  is obtained within a 60 ns time interval with a counting rate of ~200 Hz per station. The whole
   experiment is triggered by a fourfold coincidence from those particle stations in a 600 ns time
    window. The corresponding rate is around 0.7 event per minute. Considering the active area of 
    the particle detector array of $7.10^{3}$ m$^2$ and the arrival direction distribution of the
     particle pancake, a value of $16.10^3$ m$^2$.sr is obtained for the acceptance, which
      corresponds to a trigger energy threshold of about $1.10^{15}$ eV.

The recognition of the radio transients is made during an offline analysis (see also
 Ref.~\cite{dal03}$^,$~\cite{dal04}$^,$~\cite{ard04}). Radio signals are first 37-70 MHz 
 numerically filtered to detect radio transient. The maximum voltage is searched in a given
  time window, correlated to the trigger time, and compared to a threshold based on the noise 
  level estimation outside this window. If the threshold condition is fulfilled, the arrival 
  time is set at the maximum voltage and the antenna is flagged 'fired'. When at least 3 antennas
   are flagged a triangulation procedure calculates the arrival direction of the radio wave using
    a plane front. To avoid fortuitous events, a cut is applied on the arrival time difference
     between the radio wave front and the particle front (obtained with particle detectors) that 
     is within 100 ns for an EAS event. Finally,  true radio-particle 
     coincidences are selected by requiring that the arrival directions obtained by
     both particles and radio signals are the same within 15 degrees.

Due to the low trigger threshold, only a fraction of those air shower events goes with significant
 radio signal. After the previous sorting, the EAS radio event rate is 0.9 per day. Assuming that 
 the acceptance of both, particle detector and antenna array, are the same, the energy threshold
  of the radio events has been estimated around $5.10^{16}$ eV. 

At the end of these analysis procedures, physical characteristics of the radio EAS events can be
 extracted. 

\section{Illustrative EAS event}
With our limited array of antennas, the number of tagged antennas per event is highly variable,
 depending on the shower energy and core position. Thus, only events falling inside the surface 
 delimited by the extremities of our antenna array can be unambiguously analyzed. The EAS event
  example shown on Fig.~\ref{fig:chp_fh} is one of those. It exhibits a 11 antennas multiplicity
   with the associated signal amplitude measured on each antenna (depicted by the circle size).
    The arrival direction has been reconstructed from both scintillator and antenna data 
    (discrepancy of 1.6 deg. between both estimations for this event) and indicates that it
     corresponds to a shower with a zenith angle of 51 degrees.

\begin{figure}[h]
\begin{minipage}[t]{.45\textwidth}
   \includegraphics[height=4.5cm]{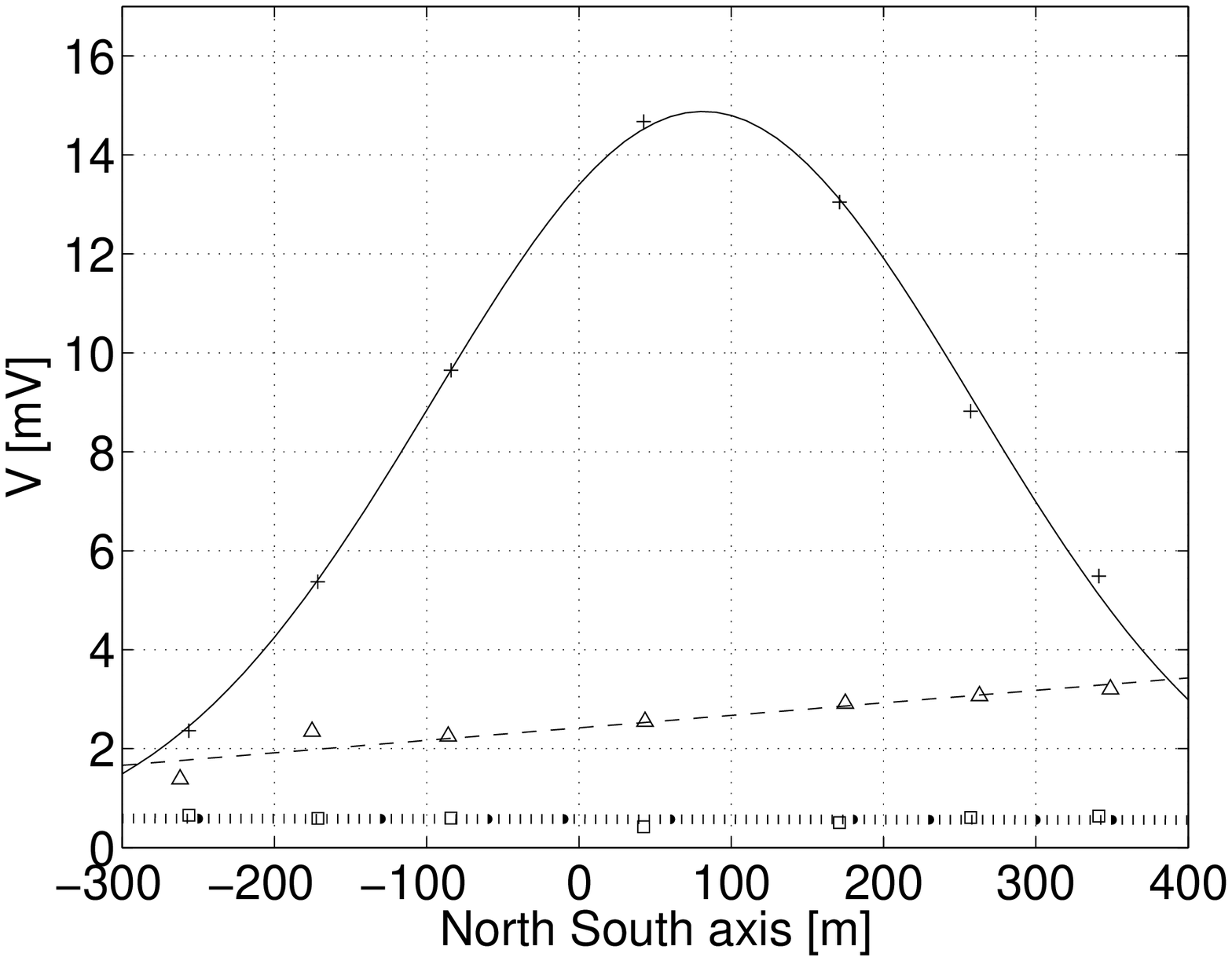}
   
\end{minipage}
\centering
\begin{minipage}[t]{.45\textwidth}
   \includegraphics[height=4.5cm]{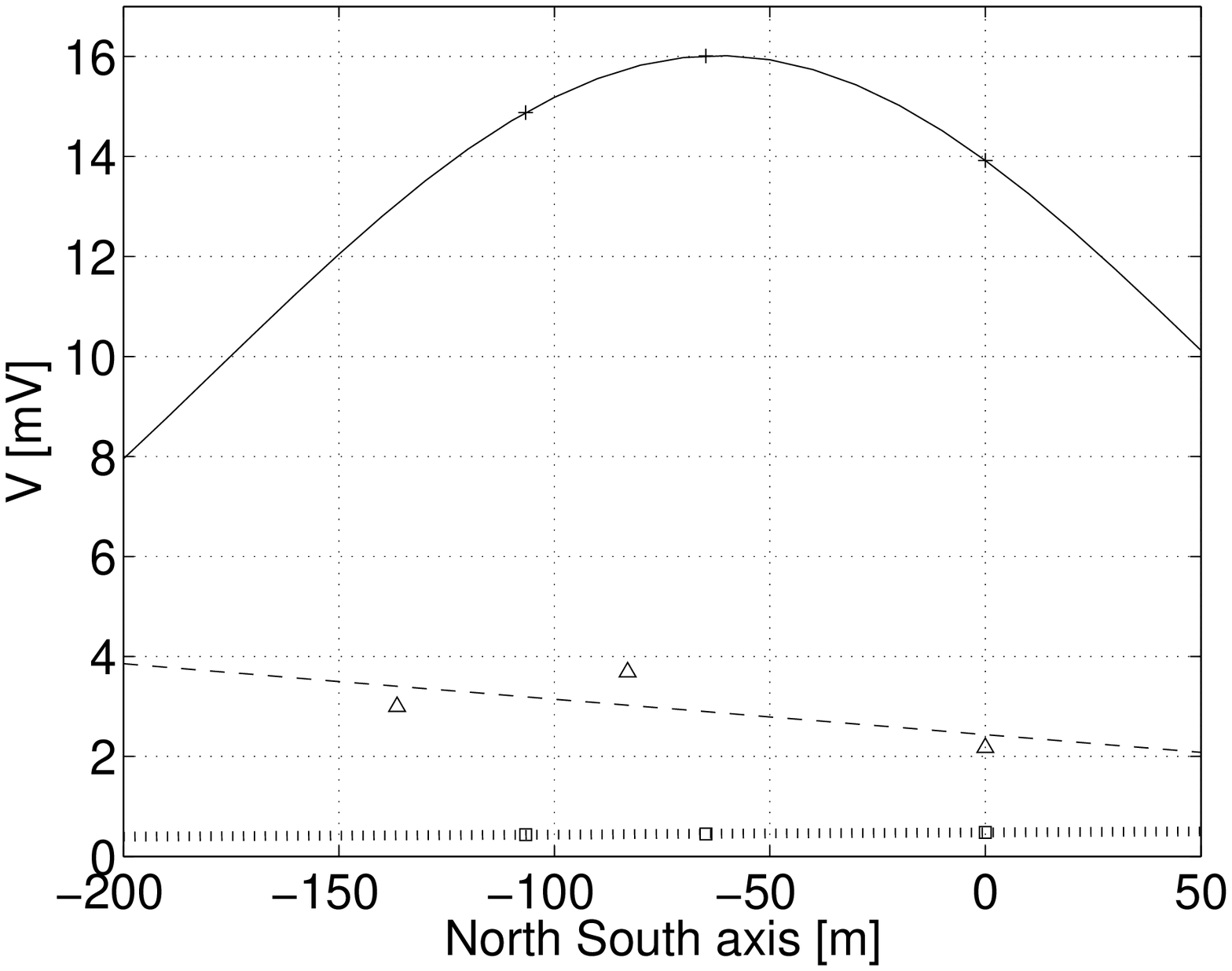}
\end{minipage}
\caption{Sampling of the electric spread. Left : Profile of the 
maximum voltage recorded on the antennas in the East-West direction. Right : Profile of
 the maximum voltage recorded on the antennas in the North-South direction. Both for an 
 EAS event (full line) and an anthropic transient (dashed line). The dot line correspond
  to the noise level of the antennas. }
\label{fig:profil}
\end{figure}

The electric field distribution for the above mentioned EAS event (full line) 
and a fortuitous one  (dashed line) are
shown Fig.~\ref{fig:profil}, for East-West (left)
 and North-South (right), antenna axis. The fortuitous event belongs to a set of events identified
  as resulting from a single polluting source (constant arrival direction from one event to 
  another) and rejected from EAS candidate status using the procedure described in
   Ref.~\cite{nim-ard}. The dotted line indicates the  noise level of our setup
    and illustrate the amount of useful signal received. Antenna responses were cross calibrated 
    and gains adjusted within a few \%.

It appears that topologies are clearly different between EAS and anthropic events. On the one side,
 the anthropic event (dashed line) presents an electric field topology with a soft linear decrease
  of the amplitude which is not 
  expected for an EAS candidate falling in the vicinity of the array.
On the other side, the EAS event (full line) shows a highly variable field amplitude depending 
on the position on the axis. This allows to estimate the projected core location using first 
barycenter calculations, then, by fitting Gaussian model. The resulting core position, similar
 with a relative location error of a few meters on each axis for both estimations, is pointed 
 by a star on Fig~\ref{fig:chp_fh}.

This margin between electric field topologies depending on its origin (EAS or anthropic) could
 constitute one decisive criterion of selection as it comes from the antenna array only and not
  from a comparison to the particle detectors. In other words, it means that a radiodetection
   of cosmic rays experiment should be able to discriminate EAS event by itself.

\begin{figure}
\begin{center}
\includegraphics[height=6.cm]{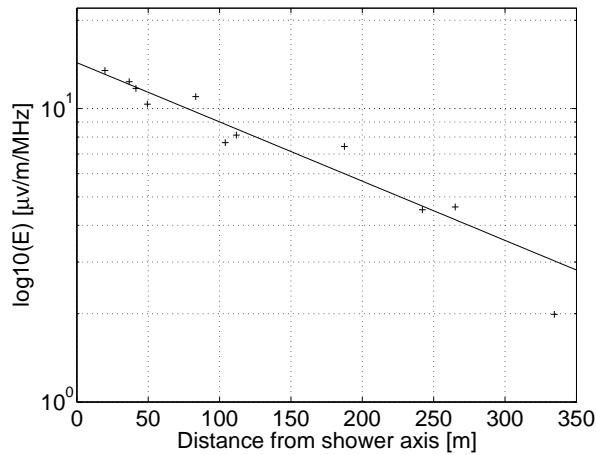}
\end{center}
\caption{The electric field measured on antennas for the presented EAS event
 as a function of the distance between the detector and the estimated shower core location.
  The full line corresponds to an exponential fit of the data.}
\label{fig:fit_chp}
\end{figure}

Once the core position and the arrival direction of the shower are known, the measured 
electric field can be plotted in a shower-based coordinate system (instead of a ground-based one)
 to look at the radio signal behavior with an increasing distance from the core such as shown
  Fig.~\ref{fig:fit_chp}. The above event fits an exponential decay given Eq.~\ref{eq:fit}~\cite{allan}.
\begin{equation}
E(d) = E_0.exp[\frac{-d}{d_0}].\
\label{eq:fit}
\end{equation}
with $d$ the distance between the shower core and the detector, $d_0$ and $E_0$ the 
fit parameters. The result gives values for those
 two parameters, $d_0$=215 m and $E_0$=14~$\mu$V/m/MHz at 37~MHz.

\section{Conclusion}

CODALEMA is already an operating tool to detect EAS associated radio emission.
It is also able to determine its arrival direction using triangulation and core
position by fitting a Gaussian profile over two axis. More data and
upgradings are needed to enhance the knowledge of the electric field
over long distances and to calibrate the experiment in energy using
additional particle detectors. Nevertheless, it is now
possible to discriminate an EAS event electric field from a fortuitous one using
only antennas and no particle detector. Considering the low trigger rate ($\leq
$ 1 evt/s) obtained during the first phase of CODALEMA ~\cite{nim-ard} where the
 system was self-triggered using a devoted antenna, this is one further step toward
  a stand-alone system that could be deployed over a large area.


\begin{thebibliography}{00}

\bibitem{agasa} N. Hayashida {\it et al}, Phys. Rev. Lett. 73, 3491 (1994); M. Takeda et al., Phys. Rev. Lett. 81, 1163 (1998).

\bibitem{fly} D.J. Bird {\it et al}, Phys. Rev. Lett. 71, 3401 (1993); Astro-phys. J. 441, 144 (1995).

\bibitem{auger} Auger Collaboration, Nucl. Instrum. Meth. A 523 (2004) 50-95 

\bibitem{casa-mia} K. Green {\it et al}, Nucl. Instrum. Meth. A498, 256 (2003).

\bibitem{cascade-grande} A. Badea {\it et al}, Proceedings of CRIS2004, Nucl. Phys. Proc. Suppl. 136, 384 (2004); astro-ph/0409319.

\bibitem{nim-ard} D. Ardouin {\it et al}, submitted to Nucl. Instrum. Meth. A, Astro-ph/0504297. 

\bibitem{ask} G.A. Askar'yan, Soviet Physics, J.E.T.P., 14, (2) 441 (1962) 

\bibitem{allan} H.R. Allan, in: Progress in elementary particle and cosmic ray physics, ed. by J.G. Wilson and S.A. Wouthuysen (North Holland, 1971), p. 169.

\bibitem{prl-ard} D. Ardouin {\it et al}, submitted to Phys. Rev. Lett., Astro-ph/0504240. 

\bibitem{huege} T. Huege and H. Falcke, astro-ph/0501580. 


\bibitem{rav04} O. Ravel {\it et al}, Proceedings of the 8th Pisa Meeting on Advanced Detectors "Frontiers Detectors for Frontier Physics", Nucl. Instr. Meth. A518, 213 (2004).

\bibitem{DAM} http://www.obs-nancay.fr/  (2005)

\bibitem{boratav} M. Boratav, {\it et al}, Proceedings of the 24th ICRC, Rome, 954(1995).

\bibitem{dal03} R. Dallier {\it et al}, \textit{SF2A 2003 Scientific Highlights}, ed. F. Combes, \textit{et al.} (EDP Sciences, 2003).

\bibitem{dal04} A. Bell\'etoile {\it et al}, \textit{SF2A 2004 Scientific Highlights}, ed. F. Combes \textit{et al.} (EDP Sciences, 2004), astro-ph/0409039.

\bibitem{ard04} D. Ardouin, {\it et al}, Proceedings of the 19th European Cosmic Ray Symposium, Florence, 2004, astro-ph/0412211.

\end{thebibliography}
\end{document}